%% file: main.tex
\documentclass[fleqn,10pt]{wlscirep}
\usepackage[utf8]{inputenc}
\usepackage[T1]{fontenc}
\usepackage{bm}

\usepackage{booktabs}
\usepackage{multirow}
\usepackage{graphicx}
\usepackage{longtable}
\usepackage{caption}
\usepackage{xurl}

\title{VisionFM: a Multi-Modal Multi-Task Vision Foundation Model for Generalist Ophthalmic Artificial Intelligence}

\author[1]{Technical Team: Jianing Qiu}
\author[1]{Hao Wei}
\author[1]{Peilun Shi}
\author[1]{Minqing Zhang}
\author[3]{Lin Li}
\author[13,14]{Benny Lo}
\author[1,*]{and Wu Yuan}
\author[2]{\\Clinical Team: Jian Wu}
\author[2]{Yunyun Sun}
\author[2]{Hanruo Liu}
\author[2]{Hongyi Liu}
\author[2]{Simeng Hou}
\author[2]{Yuyang Zhao}
\author[2]{Xuehui Shi}
\author[4]{Junfang Xian}
\author[4]{Xiaoxia Qu}
\author[2]{Sirui Zhu}
\author[2]{Lijie Pan}
\author[2]{Xiaoniao Chen}
\author[2]{Xiaojia Zhang}
\author[5]{Shuai Jiang}
\author[5]{Kebing Wang}
\author[6]{Chenlong Yang}
\author[7]{Mingqiang Chen}
\author[8]{Sujie Fan}
\author[8]{Jianhua Hu}
\author[8]{Aiguo Lv}
\author[8]{Hui Miao}
\author[8]{Li Guo}
\author[8]{Shujun Zhang}
\author[9]{Cheng Pei}
\author[9]{Xiaojuan Fan}
\author[9]{Jianqin Lei}
\author[9]{Ting Wei}
\author[10]{Junguo Duan}
\author[10]{Chun Liu}
\author[11]{Xiaobo Xia}
\author[11]{Siqi Xiong}
\author[12]{Junhong Li}

\author[15,18]{Yih Chung Tham}
\author[16,17,18]{Tien Yin Wong}
\author[2,*]{Ningli Wang}

\affil[1]{Department of Biomedical Engineering, The Chinese University of Hong Kong, Hong Kong SAR}
\affil[2]{Beijing Tongren Eye Center, Beijing Tongren Hospital, Capital Medical University, Beijing, China}
\affil[3]{Department of Informatics, King’s College London, London, U.K.}
\affil[4]{Department of Radiology, Capital Medical University, Beijing, China}
\affil[5]{Intelligent Vision Plus Technology Co., Ltd., Shenzhen, China}
\affil[6]{Center for Precision Neurosurgery and Oncology, Peking University Health Science Center, Beijing, China}
\affil[7]{SightAI Technology Pte. Ltd., Singapore}
\affil[8]{Handan City Eye Hospital, Handan, China}
\affil[9]{Department of Ophthalmology, First Affiliated Hospital of Xi'an Jiaotong University, Xi'an, China}
\affil[10]{School of Ophthalmology, Chengdu University of Traditional Chinese Medicine, Chengdu, China}
\affil[11]{Eye Center of Xiangya Hospital, Central South University, Changsha, China}
\affil[12]{Shanxi Eye Hospital Affiliated to Shanxi Medical University, Taiyuan, China}
\affil[13]{Precision Robotics (Hong Kong) Co., Ltd., Hong Kong SAR}
\affil[14]{Faculty of Medicine, Imperial College London, London, U.K.}
\affil[15]{Centre for Innovation and Precision Eye Health \& Department of Ophthalmology, Yong Loo Lin School of Medicine, National University of Singapore, Singapore}
\affil[16]{Tsinghua Medicine, Tsinghua University, Beijing, China}
\affil[17]{Beijing Tsinghua Changgang Hospital, Beijing, China}
\affil[18]{Singapore Eye Research Institute, Singapore National Eye Centre, Singapore}

\affil[*]{E-mail: wningli@vip.163.com, wyuan@cuhk.edu.hk}

\begin{abstract}

We present VisionFM, a foundation model pre-trained with 3.4 million ophthalmic images from 560,457 individuals, covering a broad range of ophthalmic diseases, modalities, imaging devices, and demography. After pre-training, VisionFM provides a foundation to foster multiple ophthalmic artificial intelligence (AI) applications, such as disease screening and diagnosis, disease prognosis, subclassification of disease phenotype (lesion, vessel, and layer segmentation, landmark detection), and systemic biomarker and disease prediction, with each application enhanced with expert-level intelligence and accuracy. The generalist intelligence of VisionFM  outperformed ophthalmologists with basic (e.g., 1-3 years of clinical experience) and intermediate (e.g., 4-8 years of clinical experience) levels in jointly diagnosing 12 common ophthalmic diseases. Evaluated on a new large-scale ophthalmic disease diagnosis benchmark database comprising 23 public and 5 private hospital datasets, as well as a new large-scale segmentation and detection benchmark database that contains 23 public and 2 private hospital datasets, VisionFM outperformed strong baseline deep neural networks. The ophthalmic image representations learned by VisionFM exhibited noteworthy explainability, and  demonstrated strong generalizability to new ophthalmic modalities, disease spectrum, and imaging devices. For example, VisionFM can accurately grade diabetic retinopathy with an AUC of 0.935 using OCTA images despite never exposed to such an imaging modality during pre-training. As a foundation model, VisionFM has a large capacity to learn from diverse ophthalmic imaging data and disparate datasets. To be commensurate with this capacity, in addition to the real data used for pre-training, we also generated and leveraged synthetic ophthalmic imaging data. Experimental results revealed that synthetic data that passed visual Turing tests, can also enhance the representation learning capability of VisionFM, leading to substantial performance gains on downstream ophthalmic AI tasks. Beyond the ophthalmic AI applications developed, validated, and demonstrated in this work, substantial further applications can be achieved in an efficient and cost-effective manner using VisionFM as the foundation. Researchers and clinicians can leverage the generalist intelligence of VisionFM to solve new clinical tasks and clinical work processes with minimal annotated samples and training, vastly accelerating the development and clinical application of future ophthalmic AI.

\end{abstract}

\begin{document}

\flushbottom
\maketitle

\thispagestyle{empty}

\input{introduction}

\input{results}

\input{discussion}


\end{document}

%% file: introduction.tex
\section*{Introduction}

\begin{figure}[!t]
\centerline{\includegraphics[width=\linewidth]{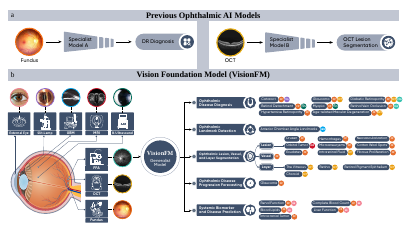}}
\caption{a. Previous ophthalmic AI models are specialized, built to be single-purpose and disease (e.g., diabetic retinopathy),  often focusing on a single modality (e.g., fundus photographs) and solving a single clinical task and process (e.g. screening). b. VisionFM is a novel AI model built to be multi-purpose, multi-disease, multi-modal, multi-task foundation model that can simultaneously approach multiple ophthalmic clinical tasks with the ability of processing multiple ophthalmic imaging modalities. VisionFM shows strong generalization to previously untrained ophthalmic modality and imaging devices, and shows robust few-shot disease diagnostic accuracy.}
\label{fig:visionfm}
\end{figure}

Vision impairment and the major ophthalmic diseases that cause visual impairment (e.g., cataract, age-related macular degeneration, glaucoma, diabetic retinopathy) are prevalent among many communities and populations. Globally, it is estimated that the number of people with a vision impairment has increased to over 2.2 billion in 2019~\cite{world2019world}. The demand for eye care services including screening and prevention, diagnosis and treatment, and community care and rehabilitation has outpaced the adequate provision and training of high-quality ophthalmologists and eye care workforce, with many countries having growing aging populations, and changes in risk factor profile affecting eye health~\cite{world2019world}. This is particularly significant in low-income countries; for example, the number of ophthalmologist per million population in low-income countries is 3.7, which is less than 5\% of the average number of ophthalmologists/population in high-income countries~\cite{resnikoff2020estimated}. This suggests a huge unmet need of delivering ophthalmic healthcare to underserved populations. To tackle global ophthalmic challenges, AI is clearly a technological solution. Over the past decade, many efforts have been made to develop ophthalmic AI to provide automated diagnoses~\cite{gulshan2016eyepacs,ting2017development,sayres2019using, dai2021deep,varadarajan2020predicting,tan2021retinal,burlina2017automated, lee2017deep,medeiros2019machine, phene2019deep, li2020development,lam2018retinal, son2020development}. While much progress has been made, the current range of AI models require much annotated data to train, and the annotations are expensive and inefficient. Current models also generally tackle single ophthalmic diseases (or a few diseases), use single modalities (e.g., fundus photographs), and clinical applications on a single task (e.g., an AI model is developed to only recognize and screen for diabetic retinopathy based on fundus images~\cite{ruamviboonsuk2022real}). These specialized single use, single modality models often have no or very limited generalization to new diseases, new imaging modalities and devices, and new clinical tasks. This has limited the application of ophthalmic AI in clinical use and cannot address the considerable unmet needs in tackling global ophthalmic challenges.

\begin{figure}[!t]
\centerline{\includegraphics[width=\linewidth]{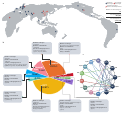}}
\caption{a. Data used for pre-training and evaluating VisionFM covers a diverse geographic locations (26 countries and regions in total). Pre-training data alone contains 3.4 million ophthalmic images from 560,457 unique individuals. b. The pre-training data of VisionFM covers eight main ophthalmic imaging modalities captured by a wide range of devices. c. Multi-modal data paired by diseases. DR: diabetic retinopathy. AMD: age-related macular degeneration. HR: hypertensive retinopathy. RVO: retinal vein occlusion. RD: retinal detachment.}
\label{fig:data_stats}
\end{figure}

Recently, AI foundation models (FMs), such as GPT-4~\cite{gpt4openai2023} and SAM~\cite{kirillov2023segment}, have emerged and has the potential to transform many research and industrial domains~\cite{qiu2023large,moor2023foundation}. FMs are models trained with a broad range of data and can be later adapted to solve a wider (rather than narrow) spectrum of tasks with their generalist intelligence, providing new opportunities to tackle the growing global ophthalmic challenges in a much more efficient, adaptable and scalable solution~\cite{bommasani2021opportunities}. The recent RETFound model~\cite{zhou2023foundation} shows such generalist intelligence in detecting diseases from retinal images. Albeit impressive, RETFound is still limited in the number of ophthalmic modalities it can process, i.e., only fundus photography and optical coherence tomography (OCT), the spectrum of clinical tasks it excels, i.e., mainly ocular disease diagnosis and prognosis, as well as prediction of systemic diseases. In diagnosing diseases, RETFound still relies on modality-specific classifiers, which is inefficient when generalizing to a broader range of ophthalmic image modalities.

Our current work aims to develop a new  foundation model for ophthalmic AI using a generalist ophthalmic image foundation model approach, which we call VisionFM\footnote{VisionFM: Vision Foundation Model. The word \textit{vision} in our model's name has two indications: one refers to human vision for which this model aims to provide generalist and accurate diagnosis and analysis, and the other refers to computer vision, based on which this model was developed.}. It was pre-trained with self-supervision using 3.4 million diverse ophthalmic imaging data (please refer to Figure~\ref{fig:data_stats}), covering a broad spectrum of ophthalmic diseases, modalities and imaging devices, and demographic data. As a foundational knowledge base of ophthalmic images, VisionFM offers a comprehensive ground to foster many downstream applications, such as disease screening and diagnosis, subclassification of disease phenotype (lesion, vessel, and layer segmentation, landmark detection), disease progression prediction and prognosis, systemic biomarker and disease prediction from ophthalmic images (please refer to Figure~\ref{fig:visionfm}). Furthermore, VisionFM has few-shot disease diagnostic ability and strong generalization to unseen imaging modality and devices. With VisionFM as a well-established ophthalmic image analytic base, the performance of many downstream applications can be enhanced, and many new applications can be implemented by sample-efficient fine-tuning to unlock its intelligence for ophthalmic image analysis.

More importantly, in real-world clinical ophthalmic practices, ophthalmologists often use multiple imaging modalities, e.g., fundus photography (FP), OCT, ultrasound biomicroscopy (UBM), slit-lamp imaging, fundus fluorescein angiography (FFA), magnetic resonance imaging (MRI), etc, to diagnose multiple diseases and perform multiple clinical tasks (detection and diagnosis and prognosis). To realize multi-modal AI, an alternative approach is to input multiple images of different modalities to the AI model, which then diagnoses the disease based on these inputs~\cite{wang2022learning}. However, this overlooks the fact that for many medical resource-poor areas, clinics and hospitals do not have a spectrum of imaging devices, and hence a patient may only have one single modality captured, e.g., fundus photography, despite the need to have diagnosis requiring multiple imaging modalities. VisionFM enables single-modal diagnostics to reach multi-modal diagnostic accuracy through multi-modal learning, and with one single modality-agnostic decoder, it can diagnose multiple ophthalmic diseases captured in different imaging modalities, be it fundus, OCT, FFA, slit-lamp or B-scan ultrasound.

As a foundation model, VisionFM has the capacity to model a large volume of ophthalmic imaging data. While it is expected that increasing the amount of real-world data will lead to better learning outcomes, it is unfeasible to curate a massive amount of real-world data, particularly for certain imaging modalities which are not commonly performed, e.g., ocular MRI. Our work hence also studies the effectiveness of synthetic ophthalmic imaging data. Our studies revealed that synthetic data generated by a large generative model that can pass visual Turing tests~\cite{salimans2016improved} conducted by a panel of clinicians can also enhance the representation learning capability of VisionFM, leading to substantial performance gains on downstream ophthalmic AI tasks. Beyond being used as additional training data, we believe high-fidelity synthetic ophthalmic image data also holds great promises in future ophthalmic education and research.

%% file: results.tex
\section*{Results}

\begin{figure}[!t]
\centerline{\includegraphics[width=\linewidth]{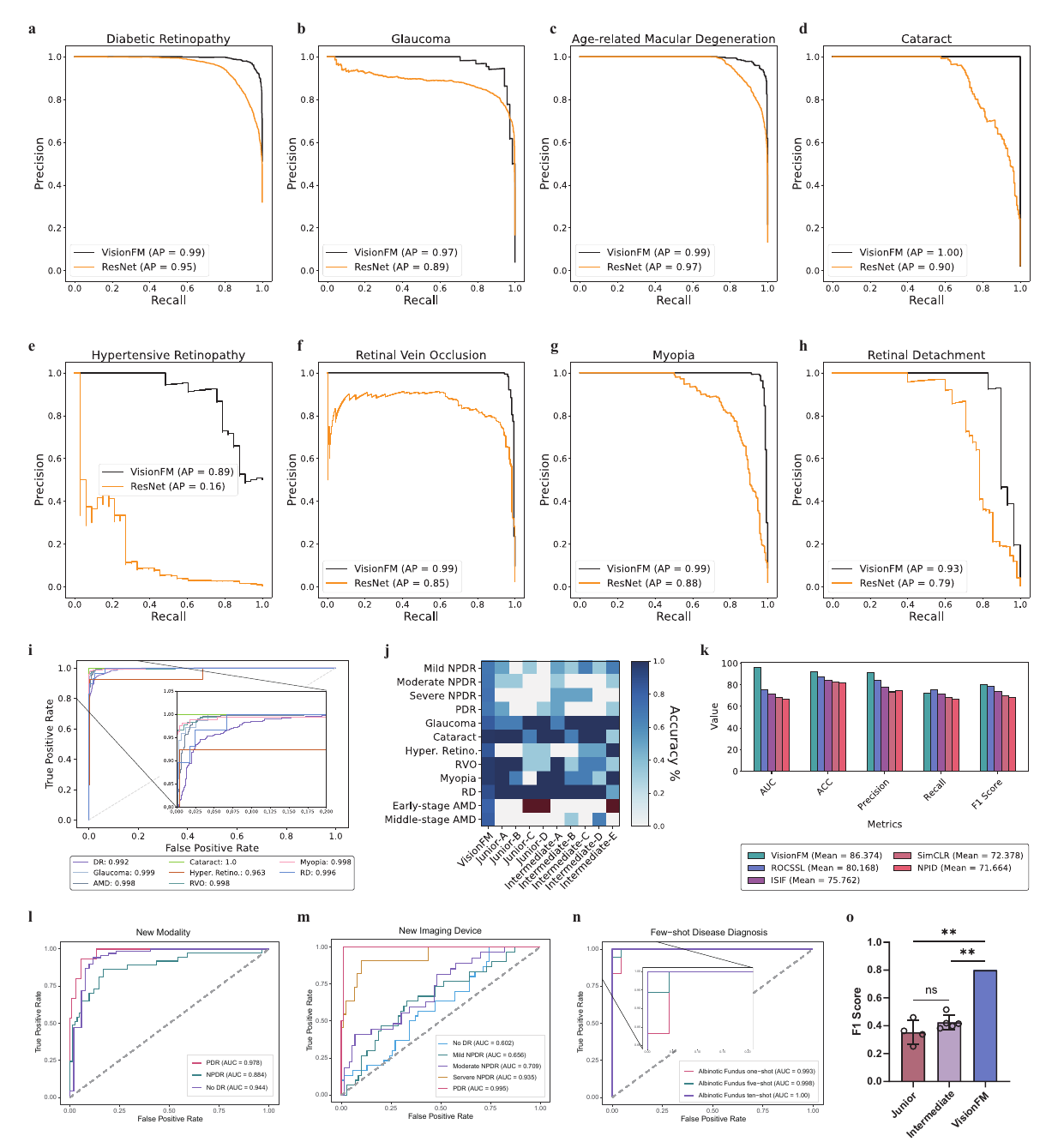}}
\caption{Ophthalmic disease diagnosis. a-h: Precision-recall curves of jointly recognizing eight common ophthalmic diseases on a large-scale benchmark that merges 23 public and 5 private datasets, covering five imaging modalities, and comparison with ResNet pre-trained on ImageNet and later fine-tuned on the same training set. i: AUC of recognizing eight ophthalmic diseases by VisionFM. j: Generalist diagnostic accuracy of VisionFM across 12 ophthalmic diseases and comparison with nine ophthalmic clinicians of different years of experience. k: Comparison with other self-supervised or unsupervised methods in recognizing AMD on the iChallenge-AMD dataset. l-n: Generalizability evaluation results. l: AUC of DR grading on a new modality (OCTA). m: AUC of DR grading on new imaging devices (ultra-wide-field fundus photography devices). n: AUC of diagnosing an underrepresented ophthalmic disease (ocular albinism) in few-shot manners. o: Average F1 score of VisionFM across 12 diseases, and comparison with the results of the two cohorts of ophthalmic clinicians.}
\label{fig:disease_diagnosis}
\end{figure}

\begin{figure}[!t]
\centerline{\includegraphics[width=\linewidth]{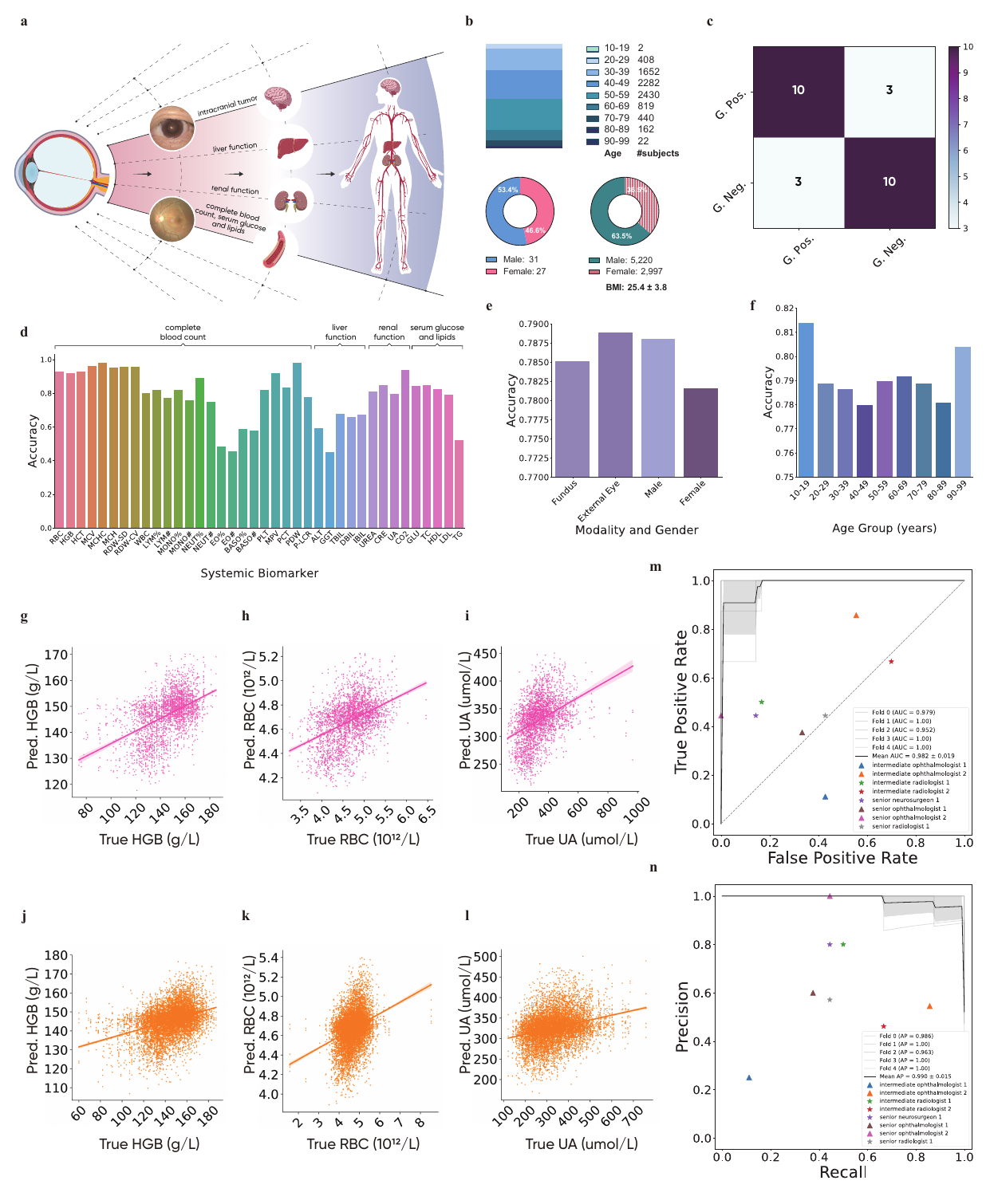}}
\caption{a: Illustration of systemic biomarker and disease prediction from ocular images. b demography of the subjects involved in systemic biomarker (top and bottom right) and disease prediction (bottom left). c: Confusion matrix of glaucoma forecasting results of VisionFM. d: Average accuracy of predicting each individual systemic biomarker from fundus and external eye images. e and f: Average accuracy of systemic biomarker prediction of different modalities and gender (e) as well as different age groups (f). g-l: Predicted and actual biomarker values of HGB, RBC, and UA (g, h, i are results of external eye images; and j, k, l are fundus images). m and n: results of predicting the presence of intracranial tumors from fundus images. m: AUC of VisionFM and comparison with clinicians. n : AP of VisionFM and comparison with clinicians.}
\label{fig:fore_sys_bio}
\end{figure}

\begin{figure}[!t]
\centerline{\includegraphics[width=\linewidth]{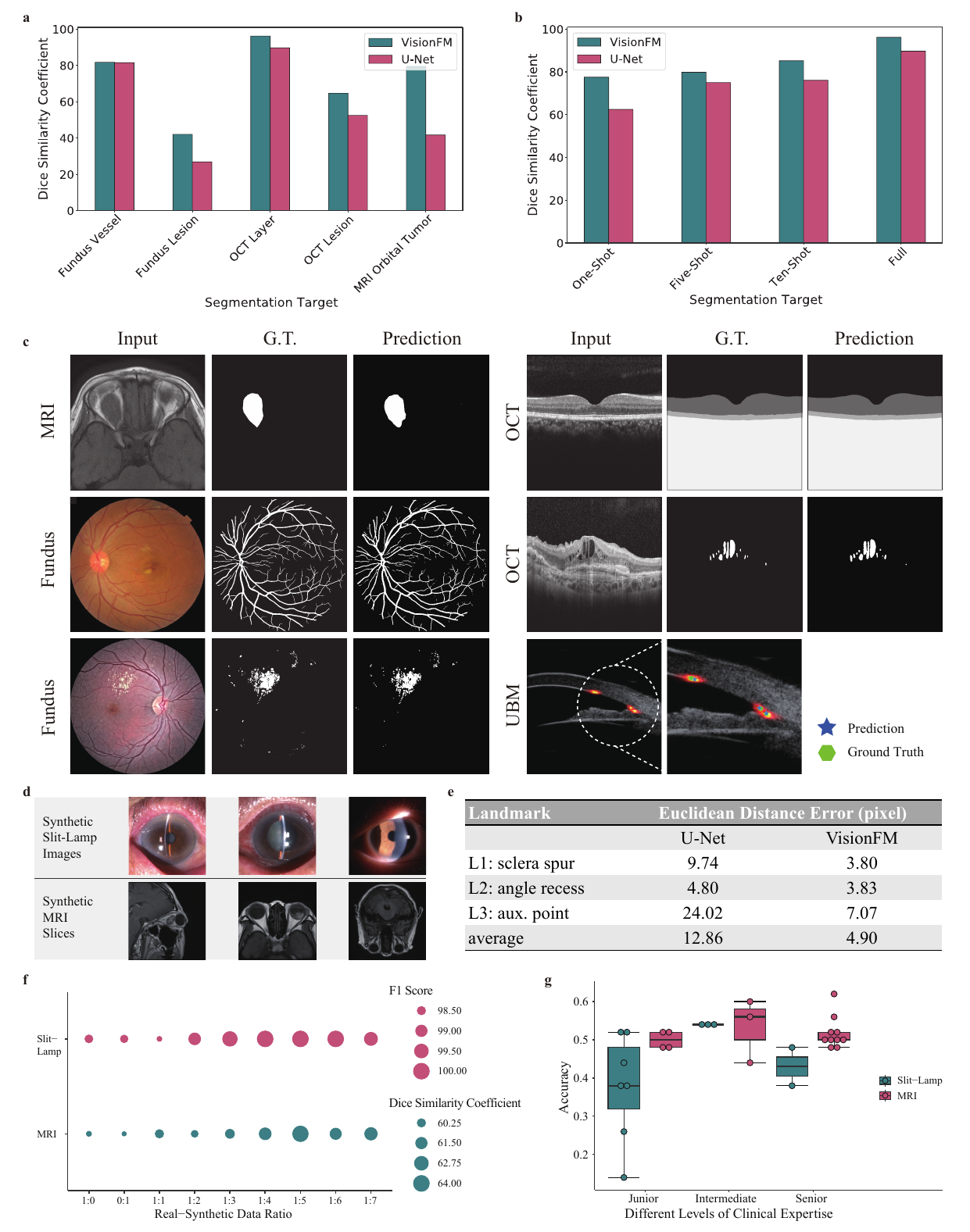}}
\caption{a: Quantitative segmentation results of VisionFM and comparison with U-Net. b: Few-shot OCT layer segmentation performance of VisionFM and comparison with U-Net. c. Qualitative examples of segmentation and landmark detection of VisionFM on different imaging modalities. d. Examples of synthetic slit-lamp and MRI images. e: Euclidean distance errors of VisionFM in detecting three landmarks in UBM images and comparison with U-Net. f: Cataract diagnostic performance of VisionFM with respect to different proportions of real and synthetic slit-lamp data, as well as orbital tumor segmentation accuracy of VisionFM with respect to different proportions of real and synthetic MRI data. g: Results of visual Turing test (a score close to 0.5 means that synthetic images are hard to be distinguished from real images) .}
\label{fig:seg_det_synthetic_results}
\end{figure}

\begin{figure}[!t]
\centerline{\includegraphics[width=\linewidth]{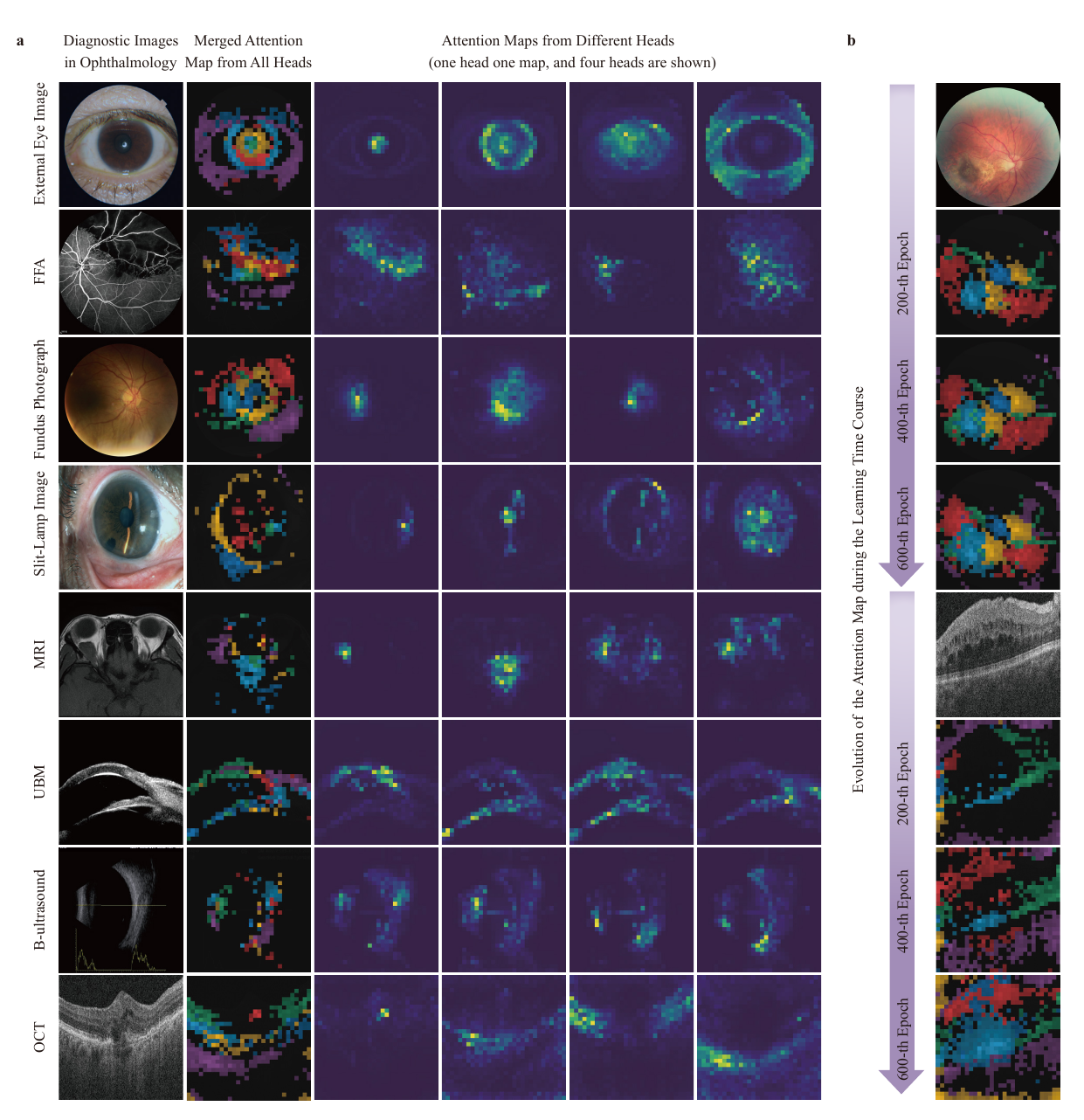}}
\caption{Explainability of the learned ophthalmic image representations of VisionFM. a: Attention maps of different heads of VisionFM on images of different modalities are visualized. b: Evolution of attention maps during the course of pre-training.}
\label{fig:attention_maps}
\end{figure}

\subsection*{Ophthalmic Disease Recognition, Grading, and Diagnostic Result Comparison with Clinicians}

After self-supervised training, VisionFM has the generalist intelligence to diagnose and grade multiple ophthalmic diseases. In order to comprehensively evaluate its performance, we merged 23 public ophthalmic disease datasets and 5 private internal datasets to curate a new benchmark that contains eight common ophthalmic diseases covering five modalities (i.e., fundus, OCT, FFA, B-ultrasound, slit-lamp). In recognizing these eight common diseases, VisionFM achieved an average area under the receiver operating characteristic curve (AUC) of 0.993 as shown in Figure~\ref{fig:disease_diagnosis}i. Note that VisionFM can recognize these diseases jointly using one single modality-agnostic decoder (i.e., one for all modalities and for recognizing eight diseases), which is more efficient than previous ophthalmic AI which has to train a separate classifier for each disease and each modality~\cite{ting2017development}, as well as the recent RETFound model~\cite{zhou2023foundation}, which still leverages modality-specific classifiers for disease diagnosis. Figure~\ref{fig:disease_diagnosis}a-h shows precision-recall curve for each individual disease, and comparison with ResNet~\cite{he2016deep} pre-trained on ImageNet and later fine-tuned on the same training set containing the eight diseases. On average, VisionFM achieved a mean average precision (AP) of 0.969, compared to 0.799 of ResNet. Its modality-agnostic implementation is efficient for generalist disease diagnosis, and a modality- and disease-specific implementation is also superior to other specialist models. For example, as shown in Figure~\ref{fig:disease_diagnosis}k, tested on the iChallenge-AMD dataset~\cite{dt4f-rt59-20}, VisionFM outperformed prior self-supervised/unsupervised methods~\cite{li2021rotation, ye2019unsupervised, chen2020simple, wu2018unsupervised} in 4 out of 5 evaluation metrics. To benchmark this generalist intelligence from a clinical experience perspective, we further compared the generalist disease diagnostic performance of VisionFM with ophthalmologists having different years of clinical experience. Nine ophthalmologists were involved, five of them are at the intermediate level (4-8 years) and four are at the junior level (1-3 years). In total, 90 ophthalmic images containing 12 ophthalmic diseases were tested, and the concerted annotations from three senior ophthalmologists (with more than 9 years of clinical experience) were used as the gold standard. 
As shown in Figure~\ref{fig:disease_diagnosis}j and Figure~\ref{fig:disease_diagnosis}o, VisionFM outperformed ophthalmic clinicians at both junior and intermediate levels. Its F1 score nearly doubled that of intermediate clinicians (82.8\% vs. 42.6\%). The results reveal that while ophthalmic clinicians may excel at one specific ophthalmic domain, e.g., cataract, in diagnosing a wide categories of ophthalmic diseases jointly, their diagnostic performance is limited, whereas VisionFM can consistently have accurate diagnosis across different ophthalmic diseases that manifest in different imaging modalities.

\subsection*{Glaucoma Progression Forecasting}

It is estimated that by 2040, there will be 111.8 million people worldwide affected by glaucoma~\cite{tham2014global}. Accurate forecast of glaucoma progression can enable early intervention, preventing under- or over-treatment to patients. We hence examined the glaucoma forecasting capability of VisionFM. Specifically, VisionFM was adapted to predict whether a subject will develop glaucoma at their next clinical visit, based on the fundus photograph taken at the current clinical visit and the time interval between the current and next visits. A single linear layer was trained on top of VisionFM for prediction. Longitudinal data of 407 subjects were used (34 subjects were diagnosed with glaucoma at their second visit). In total, 381 subjects were sampled to train the linear layer with VisionFM acting as an ophthalmic image feature extractor, and 26 subjects (half of them were diagnosed with glaucoma at their second visit) were retained for testing. The average time interval between the two visits is 1035.4 days in the training set and  935.9 days in the test set. Given its asymptomatic and chronic nature, early glaucomatous changes are challenging to identify, but nevertheless, VisionFM was still able to achieve a F1 score of 76.9\% in predicting glaucoma progression, as shown in Figure~\ref{fig:fore_sys_bio}c.

\subsection*{Systemic Biomarker and Disease Prediction from Ophthalmic Images}

Eyes are an important window to peeking into the general state of health of a person, and many ophthalmic imaging modalities manifest the signs that link to the health conditions of different parts of the body (e.g., the blood vessels captured in fundus photographs may indicate cardiovascular health~\cite{poplin2018prediction}). Ophthalmic imaging hence could provide a non-invasive way to examine the general state of health of a person~\cite{rim2020prediction,babenko2023deep}. Based on VisionFM, we conducted an exploratory study to estimate multiple systemic biomarkers from fundus and external eye images. In total, retrospective data of 8,217 subjects was used. Each subject has paired ophthalmic images and gold standard blood sample test results, and both were taken on the same half day in a screening. A multilayer perceptron (MLP) was trained on top of VisionFM to estimate 38 systemic biomarkers from ophthalmic images, and 80\% of the subjects were used for training the MLP and the rest 20\% were retained for testing. In estimating 38 different systemic biomarkers associated with complete blood count, liver function, renal function, and serum glucose and lipids, an average accuracy of $78.6\% \pm 15.2\%$ was achieved as shown in Figure~\ref{fig:fore_sys_bio}d. Notably, the estimation accuracy for renal function biomarkers is $84.8\% \pm 5.6\%$, which suggests renal and ocular health may have close association. Previous study also revealed that from retinal images, chronic kidney disease could be identified~\cite{sabanayagam2020deep}. The average accuracy of complete blood count (24 biomarkers) is $81.7\% \pm 15.1\%$, and the results revealed that for certain biomarkers, especially those related to red blood cells, the estimation accuracy can be extremely high, e.g., 98.1\% for estimating mean corpuscular hemoglobin concentration (MCHC), which we hypothesize is because ocular photographs capture details of blood vessels such as their structure, color, and diameter that may reflect health conditions pertaining to blood cells. For estimating serum glucose and lipids (five biomarkers), an average accuracy of $76.5\% \pm 12.4\%$ was achieved, in which the accuracy of estimating total cholesterol (TC) can reach to 84.7\%, and 84.4\% for serum glucose. The results also suggest that it is possible to estimate liver function biomarkers from fundus and external eye images, but the relatively low estimation accuracy (an average accuracy of $61.0\% \pm 8.5\%$ was achieved in estimating five liver function biomarkers) may suggest that liver function biomarkers may not have clear manifestation in fundus and external eye images. Figure~\ref{fig:fore_sys_bio}e and Figure~\ref{fig:fore_sys_bio}f show the average accuracy of each modality, gender, and age group.

The presence of an intracranial tumor may affect a person's vision, for example, by causing papilledema or exerting pressure on the optic nerve. Identifying these changes in ocular images can raise the awareness of further investigations. Hence, we trained a linear layer on top of VisionFM to predict the presence of intracranial tumors from fundus images. Specifically, we used retrospective data of 58 subjects, among whom 35 subjects were diagnosed with an intracranial tumor. 5-fold cross validation was conducted to evaluate the intracranial tumor prediction capability of VisionFM. As shown in Figure~\ref{fig:fore_sys_bio}m and Figure~\ref{fig:fore_sys_bio}n, VisionFM achieved an average AUC of 0.982 and AP of 0.990, vastly exceeding intermediate-level (4-8 years of clinical experience) and senior-level (at least 9 years of experience) clinicians in predicting intracranial tumors from fundus images.

\subsection*{Generalization to New Modality, Disease, and Imaging Device}

Apart from accuracy, whether an AI model has the capability of generalizing to new devices, and potentially new diseases and modalities is an important factor to determine whether it is robust and scalable in real-world deployment. We hence examined the generalizability of VisionFM from three perspectives: 1) new ophthalmic modality (we examined how well it can perform DR grading on optical coherence tomography angiography (OCTA) images despite not seeing such a modality during pre-training); 2) new imaging device (we examined how well it can perform DR grading on images captured by ultra-wide-field fundus photography devices despite not seeing any such photographs during pre-training); 3) underrepresented ophthalmic diseases (we examined how well it can diagnose diseases underrepresented in the pre-training data in one-, five-, and ten-shot manners). All tests were conducted by simply training a single linear layer on top of VisionFM without updating its weights (linear probing). Overall, VisionFM demonstrated decent generalization across these three aspects as shown in Figure~\ref{fig:disease_diagnosis}l-n. Notably, an average AUC of 0.935 was achieved in grading DR with OCTA images on the public DRAC dataset~\cite{qian2023drac}, despite the fact that it never saw OCTA data during pre-training.  VisionFM also demonstrated decent generalizability to completely new imaging devices, and achieved an average AUC of 0.779 in grading DR on the public DeepDRiD dataset~\cite{liu2022deepdrid}, containing fundus images captured by ultra-wide-field fundus imaging devices. In recognizing ocular albinism (OA) that is underrepresented in the pre-training data, with one annotated sample as illustration (one-shot), VisionFM can achieve an AUC of 0.993, and with five samples (five-shot), the AUC increases to 0.998, and with ten samples (ten-shot), it can already correctly identify all test data in an internal dataset comprising 18 OA and 23 non-OA fundus images.

\subsection*{Ophthalmic Lesion, Vessel, Layer Segmentation, and Landmark Detection}

Accurately segmenting ophthalmic lesions, vessels, and layers can support ophthalmologists to make better diagnostic decisions. VisionFM demonstrates accurate segmentation accuracy on ophthalmic imaging of different modalities, and targets of different presentations and topology. To thoroughly evaluate its segmentation performance, we curated an ophthalmic segmentation benchmark by merging 23 public datasets and a private internal MRI dataset. Figure~\ref{fig:seg_det_synthetic_results}a summarizes the quantitative segmentation results. In segmenting blood vessels in fundus photographs, VisionFM achieved a Dice similarity coefficient of 81.75\%, and a Dice of 42.07\% in segmenting lesions, compared to 81.50\% and 26.87\% of U-Net~\cite{ronneberger2015u} in segmenting vessels and lesions in fundus photographs. In segmenting layers in OCT images, it achieved a Dice of 96.18\%, and a Dice of 64.65\% in OCT lesion segmentation, compared to 89.70\% and 52.48\% of U-Net in these two tasks. We also propose two novel tasks, i.e., orbital tumor segmentation from MRI (in which VisionFM achieved a Dice score of 79.49\%, compared to 41.69\% of U-Net), and UBM landmark detection to facilitate the quantitative measurements of the anterior chamber angle (ACA). We propose to detect three key landmarks on UBM images for quantitatively measuring ACA, which can provide objective and accurate diagnoses for angle closure glaucoma. Specifically, landmark one is sclera spur; landmark two is angle recess; and landmark three is an auxiliary point, which is the intersecting point of the tangent of landmark one and posterior cornea. In detecting three landmarks in UBM images, VisionFM achieved an average Euclidean distance error of 4.90 pixels between the predicted and ground truth landmark points (3.80 pixels for landmark one; 3.83 pixels for landmark two, and 7.07 pixels for landmark three), which is much lower compared to the average error of 12.86 pixels achieved by U-Net as shown in Figure~\ref{fig:seg_det_synthetic_results}e. Few-shot segmentation capability of VisionFM has also been examined. As shown in Figure~\ref{fig:seg_det_synthetic_results}b, the one-shot segmentation capability of VisionFM (only one annotated example was used to train the segmentation decoder of VisionFM) has already surpassed the ten-shot segmentation (ten annotated examples were used) of U-Net, Dice of 77.54\% vs. 76.07\%, in segmenting OCT layers. Figure~\ref{fig:seg_det_synthetic_results}c visualizes some qualitative segmentation and detection results of VisionFM.

\subsection*{Effectiveness of Synthetic Ophthalmic Imaging Data}

As the amount of slit-lamp and MRI data is much less than those of other modalities, to enhance and robustify VisionFM's capability of learning slit-lamp and MRI representations, two large generative models were trained to generate synthetic slit-lamp and MRI data. Figure~\ref{fig:seg_det_synthetic_results}d shows some generated slit-lamp and MRI images. The generated synthetic data was randomly sampled and examined by a panel of ophthalmologists, neurosurgeons, and radiologists using visual Turing tests (i.e., distinguishing an image as real or synthetic, a score closing to 50\% means the synthetic images are of high-fidelity, and hard to be differentiated from real images). Overall, synthetic slit-lamp data obtained a score of $42.7\% \pm 12.1\%$, and MRI data a score of $51.6\% \pm 4.5\%$ as shown in Figure~\ref{fig:seg_det_synthetic_results}g. 

We then conducted thorough ablation studies to quantify how synthetic data can contribute to the learning outcomes of VisionFM. Figure~\ref{fig:seg_det_synthetic_results}f compares the performance of VisionFM encoders that were trained with real data, synthetic data, and combination of real and synthetic data with different proportions. The results revealed that a proper ratio between the amount of real and synthetic data can lead to substantial performance gains, which, for both slit-lamp and MRI, reached to the maximum at a ratio of 1:5 (real:synthetic). It is also worth noting that training with only real data or only synthetic data can converge to similar accuracy, showing the potential of replacing real data with synthetic ones to encourage multi-center data sharing in the future. In addition, while gradually adding high-fidelity synthetic data will generally improve the performance, using excessive synthetic data can start to decrease the performance gains, which we hypothesize that an effective equilibrium between real and synthetic data distributions is important while harnessing the benefits of synthetic data, and the excess of synthetic data might break this equilibrium, resulting in sub-optimal performance gains.

\subsection*{Explainability of Learned Ophthalmic Imaging Representations of VisionFM}

The generalist intelligence of VisionFM on multiple ophthalmic downstream tasks largely depends on its strong image representation learning capability. To better understand how VisionFM interprets ophthalmic images, we visualize the attention maps from different heads of VisionFM on images of different ophthalmic modalities. As shown in Figure~\ref{fig:attention_maps}a, different heads will attend to different regions of an ophthalmic image, and the merged attention map from all heads mostly focus on the foreground region, and targets of interest in an ophthalmic image. For example, in analyzing external-eye images, the heads of VisionFM will attend to the pupil, iris, sclera, and also surrounding regions of the eye, and for fundus and FFA images, the heads will focus on optic disc and cup, macula, blood vessels, and lesions (e.g., the neovascularization in fundus). Notably, in analyzing slit-lamp images, VisionFM will pay particular attention to the slit-lamp light. For UBM images, VisionFM will attend to sclera spur and angle recess, and for OCT images, layer structures, and lesions if present are VisionFM's major focus. We also visualize the evolution of the attention maps during the time course of pre-training. As shown in Figure~\ref{fig:attention_maps}b, as learning continues, VisionFM will gradually attend to more semantically meaningful regions of an ophthalmic image (e.g., it gradually learns to focus on lesions when pre-training continues).

%% file: discussion.tex
\section*{Discussion}

As AI develops into mainstream medical technology, its ability should be expanded to solve a wider and more complex range of clinical tasks to meet the vast clinical and public health needs of global diseases and challenges. Here, we present VisionFM, a novel comprehensive ophthalmic image foundation model with generalist intelligence, demonstrating consistent state-of-the-art accuracy across various ophthalmic devices, modalities, applications and clinical tasks. VisionFM can diagnose ophthalmic diseases in a modality-agnostic manner, enabling one single decoder to diagnose multiple diseases present in different modalities. To the best of our knowledge, \textbf{VisionFM is the first model that enables this modality-agnostic diagnosis.} Notably, it outperformed basic and intermediate-level ophthalmologists in diagnosing 12 common ophthalmic diseases (e.g., different stages of DR and AMD) captured in five modalities. VisionFM also enables systemic biomarkers to be predicted using this efficient modality-agnostic way from fundus and external eye images. In addition, \textbf{it is the first foundation model that supports segmentation (e.g., ocular lesion, vessel, layer, and orbital tumor) and detection (e.g., ACA landmark), beyond recognition (e.g., multiple ophthalmic diseases) and prediction (e.g., glaucoma progression).}

\textbf{VisionFM shows multi-faceted generalization capabilities.} For example, it can generalize to OCTA images for DR grading despite never being exposued to such a modality during pre-training, and to new imaging devices (e.g., ultra-wide-field fundus imaging) that were not used to capture data for pre-training. With sample-efficient fine-tuning (e.g., one-, five-, and ten-shot learning), VisionFM can be adapted to diagnose a new disease (e.g., ocular albinism) with high accuracy. To further examine the generalist intelligence of VisionFM, we introduced and benchmarked a series of new ophthalmic tasks in this work, such as orbital tumor segmentation, UBM landmark detection, and intracranial tumor prediction. Results from our systemic biomarker predication study present new evidence that ocular conditions captured in fundus and external eye images can reflect other body conditions, such as renal function, complete blood count, blood lipids, but the correlation between ocular and liver health requires further investigation as the results suggest it is comparatively hard to accurately estimate liver function biomarkers from fundus and external eye images, and a previous study also found that while severe hepatobiliary diseases could be identified from ocular images, mild ones are hard to be identified~\cite{xiao2021screening}. \textbf{While it is relatively hard for ophthalmologists and radiologists to infer the presence of an intracranial tumor from a fundus image, VisionFM can make accurate prediction directly from fundus images, and it is the first AI model that has been validated for this task}. Finally, to bridge the gap between the learning capacity and the availability of high-quality real-world data, we proposed to use synthetic data to strengthen the learning of ophthalmic image features. We demonstrated that with a proper proportion of real to synthetic data, the performance of VisionFM can be enhanced, and such enhancement may also emerge in other foundation models while learning with synthetic data. The generated synthetic data holds prospects for international collaborative projects with privacy-preserved data sharing~\cite{kumar2022evaluation} and ophthalmic education in the future. 

The validated versatility and accuracy of VisionFM on multiple ophthalmic applications provide many opportunities for clinical translation, delivering perceivable benefits to both ophthalmologists and patients. We highlight that the introduction of VisionFM, as a foundational ophthalmic image analytic model, can foster an ecosystem for future ophthalmic AI applications, accelerating their development and also improving their performance. Despite being trained as a generalist ophthalmic AI, the diagnostic and image analytic abilities of VisionFM actually have reached to specialist accuracy in many domains such as disease diagnosis, which is particularly promising for large-scale ophthalmic disease screening in the public and diagnostic assistance in outpatient clinics, increasing the productivity of ophthalmologists and reducing the administrative expenditure at large. VisionFM is also particularly useful for regions and countries having a low ophthalmologist density, where people have limited access to high-quality diagnoses and eye care services. The glaucoma forecasting capability can be utilized to assist clinical decision-making, preventing both over- and under-treatment for patients. Its ability to predict systemic biomarkers relating to general state of health such as renal function from fundus and external eye images, which at present are estimated from blood samples, shows great prospects for people to examine those biomarkers at home in a non-invasive way in the future, as they can just take a picture of one of their eyes and query VisionFM for estimation. VisionFM can also be used in education, acting as a senior ophthalmologist to train junior ophthalmic practitioners, with its wide spectrum of knowledge in ophthalmic imaging and disease diagnostics. Furthermore, it can be integrated with a large language model (LLM) to generate diagnostic reports, closing the loop of ophthalmic disease diagnoses.

While the generalist ability of VisionFM has been extensively evaluated, its limitations should also be acknowledged. First, although the pre-training data of VisionFM covers a wide range of demographic diversity, the ethnicity might be biased towards the Chinese population as the amount of internal pre-training data outweighs external public data. Although in extensive evaluation, VisionFM demonstrates remarkable accuracy on benchmarks that have diverse ethnicity, as well as generalization to ophthalmic data captured by new imaging devices, we conjecture a more balanced and inclusive pre-training dataset may further improve the accuracy and generalizability of VisionFM, potentially enhancing the diagnoses of some rare diseases. Second, although VisionFM unifies and tackles multiple ophthalmic tasks that cover eight modalities, ophthalmology inherently is complex and there are many more modalities, diseases (especially rare diseases) and clinically useful tasks that VisionFM has yet covered. However, as the structure of VisionFM is designed to be flexible, new modalities, diseases, and tasks can be easily incorporated into the current structure with new encoders and sample-efficient fine-tuning of decoders or even simple linear probing. Hence, the clinical utility of VisionFM can be extended and verified without much hurdle when more data becomes available, broadening future benefits to the ophthalmic society and patients.

Looking forward, we envision VisionFM would inspire efforts to herald new breakthroughs in three exciting areas. The first is disease prognosis and progression prediction. Accurate prediction of disease prognosis and progression would bring tremendous benefits to patients. Prediction from AI, however, should be strengthened, and the outcome should be factually grounded such as early-stage disease manifestation in the captured imaging. While VisionFM has shown promising accuracy of predicting glaucoma progression, modalities such as medical history have not been leveraged, which combined with imaging, could potentially enhance the prediction accuracy. Other ophthalmic disease prognosis prediction could potentially be implemented with a similar paradigm using their longitudinal data for model adaptation, and the multi-modal nature of foundation models can potentially improve the prediction reliability. Second, ophthalmic imaging is comparatively inexpensive and prevalent, and hence it is encouraging to examine and study different aspects concerning the health of a person using their eyes as a window~\cite{rim2020prediction,babenko2023deep}. While we present new evidence that from ophthalmic images, systemic biomarkers, such as renal function and complete blood count, and systemic diseases such as intracranial tumors, can be inferred, the information carried in ophthalmic data may be much beyond that. Connecting ophthalmic data with data from other parts of the body may reveal many unknown associations between diseases and treatment outcomes, and the advances in foundation models may accelerate such revelation and discovery. Third, as the curation of large-scale ophthalmic data is prohibitively costly, and the amount often could not meet the learning capacity of a foundation model, we envision with advances in generative AI, synthetic data will play an increasingly important role in the future to overcome the data scarcity, and the generation will be favored to condition on diagnostic and demographic information (e.g., generating a fundus image of a 70-year-old Chinese man with PDR) to generate the right data for the model, accelerating its learning and improving its generalization.

In conclusion, VisionFM, as a multi-modal multi-task generalist foundation model, lays the groundwork and shows the prospect of ushering in a new future of ophthalmic AI. By being able to tackle eight clinical ophthalmic imaging modalities and a multitude of clinical tasks, it helps accelerate ophthalmic clinical decision-makings and management. Functionality of VisionFM can be further extended beyond the current tasks by self-supervised pre-training of new imaging modalities and supervised fine-tuning of new clinical tasks, with the potential of addressing diverse, global ophthalmic diseases and different clinical challenges. The development, validation, and release of VisionFM will help improve the accuracy, efficiency, and accessibility of AI-enabled ophthalmic diagnostics.
